\documentclass[12pt]{article} 
\usepackage[margin=1in]{geometry} 
\usepackage{amsmath,amsthm,amssymb} 
\usepackage{graphicx} 
\usepackage{float} 
\usepackage{multirow} 
\usepackage{booktabs} 

\usepackage[titletoc,title]{appendix}
\usepackage{etoolbox}
\patchcmd{\appendices}{\quad}{: }{}{}

\usepackage{titlesec}
\titlelabel{\thetitle.\quad}

\usepackage[style=authoryear,giveninits = true,maxbibnames = 99,backend=bibtex]{biblatex}
\DeclareNameAlias{sortname}{last-first} 
\DeclareFieldFormat*{title}{#1} 
\DeclareFieldFormat[article]{pages}{#1} 
\DeclareFieldFormat[article]{volume}{\mkbibemph{#1}} 
\renewcommand{\labelnamepunct}{\space} 
\renewbibmacro{in:}{\ifentrytype{article}{}{\printtext{\bibstring{in}\intitlepunct}}} 

\renewbibmacro*{journal+issuetitle}{%
	\usebibmacro{journal}%
	\setunit*{\addcomma\space}%
	\iffieldundef{series}
	{}
	{\newunit
		\printfield{series}%
		\setunit{\addspace}}%
	\usebibmacro{volume+number+eid}%
	\setunit{\addspace}%
	\usebibmacro{issue+date}%
	\setunit{\addcolon\space}%
	\usebibmacro{issue}%
	\newunit}

\renewbibmacro*{volume+number+eid}{%
	\printfield{volume}%
	\printfield{number}%
	\setunit{\addcomma\space}%
	\printfield{eid}}
\DeclareFieldFormat[article]{number}{\mkbibparens{#1}}

\bibliography{bib} 

\newcommand{\dhi}{\hat{d}_i}
\newcommand{\mhi}{\hat{m}_i}
\newcommand{\ahi}{\hat{a}_i}
\newcommand{\bhi}{\hat{b}_i}
\newcommand{\tauhi}{\hat{\tau}_i}
\newcommand{\E}{\text{E}}
\newcommand{\Var}{\text{Var}}
\newcommand{\Cov}{\text{Cov}}
\newcommand{\bparens}[1]{\left(#1\right)}
\newcommand{\bsq}[1]{\left[#1\right]}
\newcommand{\tauh}[1]{\hat{\tau}_{#1}}

\newcommand{\dhj}{\hat{d}_{j}}

\begin{document}
\title{The P-LOOP Estimator: Covariate Adjustment for Paired Experiments}
\author{
	Edward Wu\thanks{Department of Statistics, University of Michigan,  Ann Arbor, MI.}\footnotemark[1]
	\and Johann A.\ Gagnon-Bartsch\footnotemark[1]
}

\maketitle

\begin{abstract}
In paired experiments, participants are grouped into pairs with similar characteristics, and one observation from each pair is randomly assigned to treatment. Because of both the pairing and the randomization, the treatment and control groups should be well balanced; however, there may still be small chance imbalances. It may be possible to improve the precision of the treatment effect estimate by adjusting for these imbalances. Building on related work for completely randomized experiments, we propose the P-LOOP (paired leave-one-out potential outcomes) estimator for paired experiments. We leave out each pair and then impute its potential outcomes using any prediction algorithm. The imputation method is flexible; for example, we could use lasso or random forests. While similar methods exist for completely randomized experiments, covariate adjustment methods in paired experiments are relatively understudied. A unique trade-off exists for paired experiments, where it can be unclear whether to factor in pair assignments when making adjustments. We address this issue in the P-LOOP estimator by automatically deciding whether to account for the pairing when imputing the potential outcomes. By addressing this trade-off, the method has the potential to improve precision over existing methods.
\end{abstract}

\section{Introduction} \label{introduction}
In randomized controlled trials, we expect the characteristics of the treatment and control groups to be similar except for the treatment itself. However, there will often be small imbalances in baseline covariates due to chance variation in treatment assignment, which can be addressed in multiple ways. One way to improve the precision of the treatment effect estimate would be to adjust for these imbalances during the analysis. Alternatively, it might be possible to balance covariates through the design of the experiment. For example, in paired experiments, participants are organized into pairs prior to treatment assignment, and then one participant in each pair is randomly assigned to treatment. Ideally, the two participants in each pair would be as similar as possible. While a paired design is often effective, it may still be helpful to make adjustments for remaining covariate imbalances. However, perhaps in part because covariate balance is addressed through experimental design, covariate adjustment methods in paired experiments are relatively understudied.

Covariate adjustment methods can be model-based or design-based (for a discussion, see \textcite{imai2009rejoinder} and \textcite{imbens2010better}). Model-based estimators have the potential to improve efficiency; however, incorrect modeling assumptions can result in bias and increased mean squared error. Design-based estimators rely only on randomization as the basis for inference, diminishing the concern of model misspecification. Hierarchical linear models (see \textcite{raudenbush2002hierarchical} and \textcite{woltman2012introduction}) are an example of a model-based approach for blocked experiments, including paired experiments. \textcite{pinheiro2000linear} and \textcite{dixon2016should} note that hierarchical linear models are a common way to analyze blocked experiments. However, the use of such models requires one to make various modeling decisions, potentially raising concerns about model misspecification. For example, \textcite{dixon2016should} notes that there is some debate as to whether block effects should be modeled as fixed or random. 

As noted above, covariate adjustments in paired experiments are relatively understudied, and design-based methods are even more so. One recent approach is presented by \textcite{fogarty2018regression}. Fogarty examines the use of regression adjustments in paired experiments under a design-based framework, building on the work of \textcite{Freedman} and \textcite{Lin}, who discuss regression adjustments in completely randomized experiments. More recently, covariate adjustment methods have been proposed for completely randomized and Bernoulli randomized experiments that involve the use of sample splitting and machine learning methods to impute potential outcomes. These include \textcite{Aronow}, \textcite{Wager}, \textcite{chernozhukov2018double}, \textcite{wu2018loop}, \textcite{spiess2017optimal}, and \textcite{rotheflexible}. Unlike the case of regression adjustments, there is not currently an analogue to these methods for paired experiments.

In this paper, we present an analogous approach to these machine learning methods for paired experiments, the P-LOOP (paired leave-one-out potential outcomes") estimator. This method is design-based; however, it also allows for the use of models to improve performance. We leave out each pair and impute their potential outcomes using information from the remaining observations. This imputation can be done with any prediction method, such as linear regression or random forests. Regardless of the imputation method, the P-LOOP estimator is unbiased and randomization is the basis for inference. In addition, one issue when making covariate adjustments is choosing which and how many covariates to use. We can address this issue in the P-LOOP estimator by choosing an imputation method that allows for automatic variable selection. \textcite{balzer2016adaptive} and \textcite{balzer2016targeted} propose the use of targeted maximum likelihood estimation in paired experiments. As noted in \textcite{moore2009covariate}, targeted maximum likelihood estimation allows for automatic variable selection when making covariate adjustments. 

The P-LOOP estimator also addresses an issue that is specific to paired experiments, which we will call the pair inclusion trade-off. In paired experiments, the performance of the estimator can suffer if we fail to properly account for the pair assignments. If the relationship between the covariates and outcome within pairs is the opposite of the relationship overall, \textit{i.e.}, a Simpson's paradox occurs, then omitting the pair assignments will hurt precision relative to the unadjusted estimator. However, in cases where the pair assignments are not predictive of the outcome, it is better to ignore the pairing. We discuss the pair inclusion trade-off further in Section \ref{imputation}. To address the trade-off, we impute two sets of potential outcomes, one in which we account for and the other where we ignore the pair assignments. Having two sets of imputed potential outcomes, we then interpolate between them by minimizing the cross validated mean squared error. By addressing this trade-off, we protect against the Simpson's paradox, but retain the potential for improvements in precision if the pairing is not informative.

Covariate adjustment methods have also been proposed for matched-pair cluster randomized trials. For example, \textcite{small2008randomization} propose a design-based estimator, while \textcite{wu2014estimation} propose a method that assumes a superpopulation.

This paper is organized as follows. In Section \ref{background}, we discuss the model and introduce notation. In Section \ref{ploop}, we present the P-LOOP estimator and derive a variance estimate. We discuss the pair inclusion trade-off further and present an imputation method to address it in Section \ref{imputation}. In Section \ref{results}, we apply the P-LOOP estimator to simulated and actual experimental data. Section \ref{discussion} concludes.

\section{Background and Notation} \label{background}
\subsection{Estimating the Average Treatment Effect} \label{notation}
In this paper, we work under the Neyman-Rubin model (see \textcite{Neyman} and \textcite{Rubin}), a non-parametric model that is often used to analyze randomized experiments. Consider a paired randomized experiment in which there are $2N$ individuals, indexed by $i = 1, 2, ..., 2N$. 
We let $T_{i} = 1$ if the participant is assigned to treatment and $T_{i} = 0$ if control. Each of the $2N$ participants has two fixed (non-random) potential outcomes, $t_{i}$ and $c_{i}$. We observe $t_{i}$ if participant $i$ is assigned to treatment and $c_{i}$ otherwise. That is, the observed outcome $Y_{i}$ for participant $i$ is  
\begin{equation}
Y_{i} = T_{i}t_{i} + (1-T_{i})c_{i}.  \nonumber
\end{equation} 
We define the individual treatment effect for each participant as $t_{i} - c_{i}$, and the average treatment effect as $$\bar{\tau}=\frac{1}{2N}\sum_{i = 1}^{2N}(t_i-c_i).$$
Consider the case where the $T_{i}$ are independent Bernoulli random variables with probability $p = 0.5$, and suppose we wish to estimate the average treatment effect. One unbiased estimate is obtained by taking the average observed outcome of the treatment group and subtracting the average observed outcome of the control group (the ``simple difference estimator"). However, for each participant, suppose we observe a $q$-dimensional vector of baseline covariates $Z_i$ prior to treatment assignment. It may be possible to use these covariates to improve the precision of the estimate over the simple difference estimator. For example, we could estimate the average treatment effect as

\begin{align}\label{loop}
\frac{1}{2N}\sum_{i = 1}^{2N} \left[2(Y_i - \hat{m}_i)T_i - 2(Y_i - \hat{m}_i)(1-T_i)\right],
\end{align}
where $\hat{m}_{i}$ is a function of $Z_{i}$. Several authors have noted an estimator of this form can be used to incorporate covariate information. For example, \textcite{Aronow} note that if $\hat{m}_{i}$ is predictive of the observed outcome $Y_{i}$, then the resulting estimate will improve over the unadjusted estimator, while \textcite{wu2018loop} suggest estimating the quantity $m_{i} = (t_{i} + c_{i})/2$. In addition, \textcite{Aronow} and \textcite{wu2018loop} note that this estimate is unbiased if $T_i$ and $\mhi$ are independent. One way to ensure this independence is by obtaining $\mhi$ through a sample splitting procedure. For example, one could leave out the $i$-th observation and calculate $\mhi$ using the remaining observations. See \textcite{Wager}, \textcite{chernozhukov2018double}, \textcite{spiess2017optimal}, and \textcite{rotheflexible} for similar estimators.

\subsection{Notation for Paired Experiments} 
We now consider the case where the participants are pair randomized. Suppose that the $2N$ participants are organized into $N$ pairs. We index the pairs by $i = 1, 2, ..., N$, each with two participants indexed by $j = 1, 2$, and the quantities defined in Section \ref{notation} are re-indexed by $i$ and $j$. For example, for participant $j$ in pair $i$, we denote the potential outcomes as $t_{ij}$ and $c_{ij}$, and define the observed outcome, treatment indicator, and covariates as $Y_{ij}$, $T_{ij}$, and $Z_{ij}$, respectively. 

For each pair, one of the two participants is randomly chosen to be assigned to treatment and the other is assigned to control. That is, $T_{i1} \sim \text{Bern}(0.5)$, and $T_{i2} = 1 - T_{i1}$. Note that the $T_{ij}$'s are not mutually independent because exactly one participant in each pair must be assigned to treatment. However, we assume the $T_{i1}$'s are mutually independent. We can therefore essentially convert our paired experiment to a Bernoulli randomized experiment by treating each pair as an experimental unit, as we describe next. 

When treating each pair as a unit, we can draw direct analogues between the notation of paired and Bernoulli randomized experiments. We denote each pair's treatment assignment by $T_i$, where $T_i = T_{i1}$. For each pair, we also observe a response variable $W_i$ and a $2q$-dimensional vector of baseline covariates $(Z_{i1}, Z_{i2})$. As with a Bernoulli randomized experiment, each pair has two potential outcomes: we observe $a_i = t_{i1} - c_{i2}$ if $T_i = 1$ and $b_i = t_{i2} - c_{i1}$ if $T_i = 0$. To differentiate these outcomes from those of the individual participants, we will refer to $a_i$ and $b_i$ as \textit{potential differences}. We define the observed difference $W_i$ as:
\begin{equation}
W_i = T_ia_i + (1-T_i)b_i.  \nonumber
\end{equation} 
We define the pair-level treatment effect $\tau_i$ as
\begin{equation}
\tau_i = \frac{(t_{i1} - c_{i1}) + (t_{i2} - c_{i2})}{2} = \frac{1}{2}(a_i+b_i) \nonumber
\end{equation}
and the average treatment effect $\bar\tau$ as
\begin{equation}
\bar{\tau} = \frac{1}{N}\sum_{i=1}^{N} \tau_i \nonumber
\end{equation}
which is our primary parameter of interest.
\section{The P-LOOP Estimator} \label{ploop}
We now present the P-LOOP estimator, which is analogous to equation (\ref{loop}), but for paired experiments. Define the quantity
\begin{align*}
d_i &= m_{i1} - m_{i2} \\
&= \frac{1}{2}(a_i - b_i),
\end{align*}
where $m_{ij} = (t_{ij} + c_{ij})/2$, and let
\begin{align*}
\hat{\tau}_i & = \left(W_i - \hat{d}_i\right)T_i + \left(W_i + \hat{d}_i\right)(1-T_i)
\end{align*}
where $\hat{d}_i$ is an estimate for $d_i$. 
Recall that for Bernoulli randomized experiments, equation (\ref{loop}) is an unbiased estimate of the average treatment effect if $\mhi$ and $T_i$ are independent. An identical argument can be used for paired experiments to show that $\hat{\tau}_i$ will be unbiased if $\dhi$ and $T_i$ are independent.

We define the P-LOOP estimator as:
\begin{align*}
\hat{\tau} &= \frac{1}{N}\sum_{i=1}^{N}\hat{\tau}_i \\
&= \frac{1}{N}\sum_{i=1}^{N}\left[\left(W_i - \hat{d}_i\right)T_i + \left(W_i + \hat{d}_i\right)(1-T_i)\right]
\end{align*}
in which we estimate $d_i$ by using a leave-one-out procedure. For each pair $i$, we drop both observations and use the remaining $N-1$ pairs to impute $a_i$ and $b_i$ using any method (such as a random forest or linear regression). We then set $\hat{d}_i = \frac{1}{2}(\hat{a}_i - \hat{b}_i)$ and repeat this procedure for all $N$ pairs to obtain $\hat{\tau}$. This leave-one-out procedure ensures that the P-LOOP estimator will be unbiased, as $\dhi$ and $T_i$ are independent. Because the P-LOOP estimator is unbiased, the mean squared error of the estimator depends only on the variance.

\subsection{Variance of the P-LOOP Estimator} \label{variance}
In Appendix \ref{truevar}, we show
\begin{align*}
\Var(\hat{\tau}_i) &= \E\bsq{(d_i - \dhi)^2} \\
&= \text{MSE}(\dhi)
\end{align*}
and thus that the variance of the P-LOOP estimator is
\begin{align*}
\Var(\hat{\tau}) 
&=  \frac{1}{N^2}\bparens{\sum_{i=1}^{N}\text{MSE}(\dhi)
	+
	\sum_{i\ne j}\gamma_{ij}}
\end{align*}
where $\gamma_{ij} = \text{Cov}(\tauhi,\tauh{j})$. We provide an unbiased estimator for $\sum_{i\ne j}\gamma_{ij}$ in Appendix \ref{negligibility}. However, in practice we suggest that the variance be estimated without this term for computational efficiency, as $\sum_{i\ne j}\gamma_{ij}$ is generally negligible (see Appendix \ref{negligibility}). For this reason, we focus on estimating $\text{MSE}(\dhi)$.

To estimate the mean squared error of $\dhi$, we express $\text{MSE}(\dhi)$ in terms of the mean squared errors of $\ahi$ and $\bhi$. In Appendix \ref{varbound}, we show that 
\begin{align*}
\frac{1}{N^2}\sum_{i = 1}^N\text{MSE}(\dhi) &\leq \frac{1}{N}\bparens{\frac{1}{4}M_a + \frac{1}{4}M_b + \frac{1}{2}\sqrt{M_aM_b}}
\end{align*}
where
\begin{align*}
M_a &= \frac{1}{N}\sum_{i = 1}^N\text{MSE}(\ahi) \\
M_b &= \frac{1}{N}\sum_{i = 1}^N\text{MSE}(\bhi).
\end{align*}
To estimate these quantities, let $\mathcal{A} = \{k: T_k = 1\}$, $\mathcal{B} = \{k: T_k = 0\}$, and $n$ be the number of elements in $\mathcal{A}$. Define the following estimates for $M_a$ and $M_b$:
\begin{align*}
\hat{M}_a &= \frac{1}{n}\sum_{i \in \mathcal{A}}(a_i - \ahi)^2 \\
\hat{M}_b &= \frac{1}{N-n}\sum_{i \in \mathcal{B}}(b_i - \bhi)^2.
\end{align*}
Having obtained estimates for $M_a$ and $M_b$, we have the following plug-in estimator for the variance of the P-LOOP estimator:
\begin{align*}
\widehat{\text{Var}}(\hat{\tau}) = \frac{1}{N}\bparens{\frac{1}{4}\hat{M}_a + \frac{1}{4}\hat{M}_b + \frac{1}{2}\sqrt{\hat{M}_a\hat{M}_b}}.
\end{align*}

Finally, note that because $\Var(\hat{\tau}_i) = \text{MSE}(\dhi)$, the performance of the estimator depends directly on how well we estimate $d_i$. One baseline approach is to set $\dhi = 0$ for all $i$, in which case $\hat{\tau}$ will exactly equal the simple difference estimator. The fact that the P-LOOP estimator reduces to the simple difference estimator in this case provides some reassurance that the leave-one-out procedure does not inherently introduce additional noise. Moreover, if we improve the estimate of $d_i$ over setting $\dhi = 0$, we will be able to improve precision beyond this baseline. Note that improving the estimate of $d_i$ is not necessarily trivial. Because we are interested in estimating the difference between $m_{i1}$ and $m_{i2}$, it does not suffice to reduce the mean squared error for the imputed potential outcomes as in the estimator of \textcite{wu2018loop}. For example, it is possible to obtain estimates of the potential outcomes that are close to the true values while having $\dhi$ of the incorrect sign. On the other hand, we could have estimates for the potential outcomes that are far from the true values that result in $\dhi$ being close to the true $d_i$.
\section{Imputation Methods of Potential Differences in Paired Experiments} \label{imputation}
We next present an imputation method to address the pair inclusion trade-off discussed in Section \ref{introduction}. We first further discuss this trade-off and then propose a method for addressing the trade-off within the P-LOOP estimator.

Note that for the P-LOOP estimator, we always drop both observations in each pair when estimating $d_i$. Thus, when we discuss the inclusion or exclusion of the paired structure when imputing potential outcomes, we refer specifically to how we treat the remaining pairs when building a prediction model. If we ignore the paired structure when imputing potential outcomes, this means we fit a model to the remaining observations as individual units. If we include the paired structure when imputing potential outcomes, this means we fit a model to the remaining observations as paired units.

\subsection{The Pair Inclusion Trade-Off} \label{motivation}
We first discuss the pair inclusion trade-off in the context of a linear model, rather than the Neyman-Rubin model, as it is perhaps easiest to understand the pair inclusion trade-off in this context.  Consider the following standard linear regression model
\begin{align*}
Y = \alpha + T\tau + P\beta  + Z\gamma  + \epsilon
\end{align*} 
where $Y$ is the observed outcome, $T$ is the treatment assignment vector, $Z$ is a covariate, and $P$ is a $2N \times N$ matrix of indicator variables that encodes the pair assignments. Suppose that there are pair effects (that is, $\beta \ne 0$), and that $Z$ is correlated with both $P$ and $T$. If we were to omit $P$ and regress $Y$ onto $T$ and $Z$, then we would bias the estimate of $\tau$. On the other hand, suppose that the pairing is not informative ($\beta = 0$). In this case, including $P$ in the regression would inflate the variance for $\hat\tau$, and it would be preferable to omit $P$ from the regression. 

This trade-off could occur with other covariate adjustment methods in paired experiments, including the P-LOOP estimator. Suppose we ignore the paired structure of the data when we train our imputation model for the potential outcomes. In this case, we model the relationship between the covariates and the outcome overall, rather than the relationship within pairs. However, if the relationship between the covariates and outcome within pairs is sufficiently different from the relationship overall, we could obtain a $\dhi$ that is far from the truth. 
One situation where this could happen is when a Simpson's paradox occurs, and the relationship within pairs between the covariates and outcome is the opposite of the overall relationship. For example, the covariates may be positively correlated with outcome overall, but negatively correlated with the outcome within pairs. 
If we ignore pair assignments when estimating $d_i$, we would infer that higher values of $Z$ are associated with higher values of $Y$. However, for a given pair, we would want higher values of $Z$ to predict lower values of $Y$. In this case, the predicted difference $d_i = m_{i1} - m_{i2}$ would be of the wrong sign, resulting in poorer performance relative to the simple difference estimator. On the other hand, if the paired structure is not predictive of the outcome, then it would be better to omit the pair assignments when imputing the potential differences.

It can be unclear whether we should account for the pair assignments when imputing the potential differences. To avoid data snooping, we propose an imputation method in this section that automatically addresses the trade-off. We first propose methods for calculating $\ahi$ and $\bhi$ that do and do not account for the pair assignments in the prediction model, producing two sets of potential differences. Having produced two estimates for each $a_i$ and $b_i$, we propose a method to automatically interpolate between them.

\subsection{Estimating $d_i$ when Pairs are not Predictive: \\ Impute Potential Outcomes Separately}
We first estimate $d_i$ without accounting for the pair assignments for the observations outside of pair $i$. To do this, we fit a model on the individual observations and then separately impute all four potential outcomes (\textit{i.e.}, $t_{i1}, c_{i1}, t_{i2},$ and $c_{i2}$) for a given pair.

More specifically, for each pair $i$, we drop both observations in the pair. We then fit a prediction algorithm on the remaining observations, ignoring the pair assignments and treating each individual as a unit. For example, we could regress $Y_{kj}$ onto $T_{kj}$ and $Z_{kj}$ for $k \ne i$. We then use this model to impute $t_{i1}, c_{i1}, t_{i2},$ and $c_{i2}$. To obtain $\hat{t}_{i1}$, we would plug in the covariates for the first observation in pair $i$ and a treatment indicator of 1. We would obtain estimates for the remaining potential outcomes similarly and set

\begin{align*}
\hat{d}_i &= \frac{1}{2}(\hat{t}_{i1}+\hat{c}_{i1}) - \frac{1}{2}(\hat{t}_{i2}+\hat{c}_{i2}).
\end{align*}

\subsection{Estimating $d_i$ when Pairs are Predictive: \\ Impute Potential Differences Directly}
Next, we propose a method that accounts for pair assignments when estimating $d_i$. Rather than imputing the potential outcomes ($t_{i1}, c_{i1}, t_{i2},$ and $c_{i2}$), we impute $a_i$ and $b_i$ directly, treating each pair as an observational unit. Recall from Section \ref{ploop} that $a_i$ and $b_i$ are analogous to the potential outcomes in an experiment with Bernoulli randomization. We can therefore apply a procedure to the paired units that is similar to the leave-one-out procedure described earlier for estimating $m_i$ in equation (\ref{loop}). For Bernoulli experiments, we would only use the control units when imputing $c_i$ and the treatment units when imputing $t_i$. However, for paired experiments $a_i$ and $b_i$ are determined by which unit is arbitrarily labeled $j=1$ and are therefore effectively interchangeable. As an example, for the $i$-th pair, we have $a_i=t_{i1}-c_{i2}$. However, if we had instead recorded the second unit in the pair first, then $a_i$ would be $t_{i2} - c_{i1}$. We can take advantage of this fact to use all observations (except those in pair $i$) when imputing each potential difference. 

In order to build a prediction model, we need to combine the covariates for each pair in some way. One way to do this would be to simply concatenate the covariate vectors for the two observations in each pair. In this case, we define $Z_i^a$ as the vector of covariates where the covariates for the treated units come first. That is, $Z_i^a = (Z_{i1},Z_{i2})$ if $T_i = 1$, and $Z_i^a = (Z_{i2},Z_{i1})$ if $T_i = 0$. Similarly, define $Z_i^b$ as the vector of covariates where the covariates for the control units come first. For example, suppose $Z_{i1} = (1,2)$, $Z_{i2} = (2,3)$, and $T_i = 0$. Then $Z_i^a = (2,3,1,2)$ and $Z_i^b = (1,2,2,3)$. In other words, $Z_i$ is the concatenated vector of covariates as it is ordered in the original data, $Z_i^a$ is the concatenated vector where the covariates for the treated unit come first, and $Z_i^b$ is the concatenated vector where the covariates for the control unit come first.

Alternatively, we may wish to transform the covariates in some way; for example, we could take the means and differences of the covariates. In this case, define $Z_i$ as $$\bparens{\frac{Z_{i1} + Z_{i2}}{2}, Z_{i1} - Z_{i2}}.$$ That is, $Z_i$ is the vector where the first $q$ entries are the averages of each covariate for the pair, and the second $q$ entries are the differences (observation 1 minus observation 2). In analogy to the concatenation example, we define $Z_i^a$ to be the means and the treatment minus control differences and $Z_i^b$ to be the means and the control minus treatment differences. 

We can now estimate $d_i$ using these combined covariates and the observed differences. We start by leaving out pair $i$. To impute $a_i$, we create a model using the observed outcomes $W_k$ (for $k \ne i$) as our response variable and the covariates $Z_k^a$ as our predictors. We then plug the covariates $Z_i$ into this model to obtain $\hat{a}_i$. To impute $b_i$, we use a similar procedure, replacing $Z_k^a$ with $Z_k^b$. Having obtained estimates $\hat{a}_i$ and $\hat{b}_i$, we set $$\hat{d}_i = \frac{1}{2}(\hat{a}_i - \hat{b}_i).$$

\subsection{Interpolating between Imputation Methods}
We have proposed two methods for imputing potential outcomes. However, we often do not know ahead of time which method will perform better. We therefore interpolate between the two methods.

For each pair $i$, we have two estimates of $a_i$: $\hat{a}_{i}^{(1)}$ and $\hat{a}_{i}^{(2)}$. We wish to obtain the value $\alpha_i$ that minimizes the distance between $a_i$ and the interpolation $\ahi = \alpha_i\hat{a}_{i}^{(1)} + (1-\alpha_i)\hat{a}_{i}^{(2)}$. However, we want $\ahi$ to be independent of $T_i$. We therefore use a leave-one-out procedure to calculate $\alpha_i$. For each $i$, we leave out pair $i$ and set $\alpha_i$ to the value that minimizes the mean squared error for the remaining observations. In other words, we have

\begin{align*}
\alpha_i= \underset{x \in [0,1]}{\text{argmin}} \sum_{k \in \mathcal{A}\backslash i} \left[a_k - \left(x\hat{a}_{k}^{(1)} + (1-x)\hat{a}_{k}^{(2)}\right)\right]^2.
\end{align*}
Taking the derivative with respect to $x$ and setting equal to 0, we have
\begin{align*}
\alpha_i &= \frac{\sum_{k \in \mathcal{A}\backslash i}(a_k - \hat{a}_{k}^{(2)})(\hat{a}_{k}^{(1)} - \hat{a}_{k}^{(2)})}{\sum_{k \in \mathcal{A}\backslash i}(\hat{a}_{k}^{(1)} - \hat{a}_{k}^{(2)})^2}.
\end{align*}
which we then restrict to be in the interval $[0,1]$. We then set our final estimate of $a_i$ to be $\ahi = \alpha_i\hat{a}_{i}^{(1)} + (1-\alpha_i)\hat{a}_{i}^{(2)}$. We use a similar procedure for $\bhi$.

\section{Results} \label{results}
We compare the performance of the P-LOOP estimator to that of other estimators. We start with a simulation to illustrate the pair inclusion trade-off. We then apply the P-LOOP estimator to data on a paired experiment involving schools in Texas.
\subsection{Simulation Results}
We compare the simple difference estimator to the P-LOOP estimator using random forests as the imputation method. Recall from earlier that we are excluding the pair assignments in our imputation method if we impute the potential outcomes ($t_{i1}, c_{i1}, t_{i2},$ and $c_{i2}$) separately, while we are including the pair assignments if we impute the potential differences ($a_i$ and $b_i$) directly. We show results using each of these imputation methods as well as the interpolation method.

Consider a hypothetical experiment where a blood pressure medication is being tested. We generate $N = 50$ pairs of twins, half of which are of ethnicity $E_i = 0$ and the other half $E_i = 1$. Next, suppose there exists a genetic mutation $Z_{ij}$. For each participant, we set $Z_{ij} \sim \text{Bernoulli}(p_k)$ for $E_{i} = k$. We set $p_1 = 0.9$ and $p_0 = 0.5$. That is, participants of ethnicity $E_i = 1$ are more likely to have the mutation. We assume that only the observed outcome $Y_{ij}$, as well as $T_i$ and $Z_{ij}$, are recorded. Suppose that ethnicity 1 has a higher baseline blood pressure than ethnicity 0 (for reasons unrelated to the mutation), but that the presence of the mutation is causally associated with lower blood pressure. We generate the outcome as:
\begin{align*}
Y_{ij} = 80 - 10T_{ij} - 5Z_{ij} + 10E_i + \epsilon_{ij}
\end{align*} 
where $\epsilon_{ij} \overset{iid}{\sim} \text{N}(0,4)$. Because participants for ethnicity $E_i = 1$ have higher baseline blood pressure, $Z_{ij}$ is positively correlated with blood pressure across all participants. Thus a Simpson's paradox occurs: overall, $Z_{ij}$ has a positive association with blood pressure, while within pairs, $Z_{ij}$ has a negative association with blood pressure. We summarize the results of this simulation in Table \ref{tab01} under the column Simpson's Paradox.

We also generate a set of potential outcomes in which the pairs contain no additional information (beyond its association with covariate $Z_{ij}$). We generate the observed outcome as:
\begin{align*}
Y_{ij} = 80 - 10T_{ij} + 5Z_{ij} + \epsilon_{ij}
\end{align*} 
where $\epsilon_{ij} \overset{iid}{\sim} \text{N}(0,4)$. In this case, $E_i$ is associated with outcome because it is associated with $Z_{ij}$, but otherwise has no effect on outcome. We summarize the results of this simulation in Table \ref{tab01} under the column Uninformative Pairs.

\begin{table}[H] 
	\caption{\label{tab01}Simulation Results}
	\centering
		\begin{tabular}{l|rr|rr}
			\toprule
			& \multicolumn{2}{c |}{\textbf{Simpson's Paradox}} & \multicolumn{2}{c}{\textbf{Uninformative Pairs}}\\
			\textbf{Method} & \textbf{True SE} & \textbf{Nominal SE} & \textbf{True SE} & \textbf{Nominal SE}\\ 
			\midrule
			Simple Difference & 0.582 & 0.585 & 0.606 & 0.604 \\
			P-LOOP (differences) & 0.389 & 0.406 & 0.392 & 0.407\\ 
			P-LOOP (outcomes) & 0.674 & 0.676 & 0.385 & 0.390\\
			P-LOOP (interpolated) & 0.387 & 0.408 & 0.389 & 0.393 \\
			\bottomrule
		\end{tabular}
	\footnotesize 
	
	\flushleft Note: Both true and nominal standard errors are estimated using 10,000 randomly generated treatment assignment vectors. The true standard error estimates have standard errors ranging from 0.002 to 0.005, while the nominal standard error estimates have standard errors ranging from 0.0001 to 0.0003.  
\end{table}
We see that in the Simpson's paradox case, imputing the potential outcomes separately (not accounting for pairs when estimating $a_i$ and $b_i$) causes inflated variance relative to the simple difference estimator, while imputing potential differences directly (accounting for pairs) results in improved performance. However, in the case where the pair assignments are uninformative, it is better to impute the potential outcomes separately. The gains in this example are relatively minor; however, we show in the next section that the improvements can be more substantial. 

\subsection{Texas Schools Data} \label{schoolsdata}
We next apply the P-LOOP estimator to data on a randomized trial involving schools in Texas, which is discussed in \textcite{pane2014effectiveness}. This trial tested the effectiveness of a computer program, the Cognitive Tutor Algebra 1 curriculum, and included 22 pairs of schools. As the outcome, we consider the passing rate of the schools on the math section of the Texas Assessment of Knowledge and Skills (TAKS) in 2008. In addition to the passing rate, we also have available as covariates the school type (middle or high school) and a pretest score, the passing rate from 2007. We estimate the average treatment effect using either just the pretest score or both the pretest score and school type as covariates. In Table \ref{tab02}, we compare the performance of P-LOOP with the simple difference estimator and the estimators of \textcite{fogarty2018regression}, which we will refer to as Regression 1 and Regression 2. Regression 1 involves the treatment minus control outcomes regressed onto the treatment minus control covariates, while Regression 2 is the same regression with the addition of the mean of the covariates in each pair. For the sake of comparison, we use linear regression as the imputation method in the P-LOOP estimator. As in the case of the simulations, we show the results imputing potential differences (accounting for pairs), imputing potential outcomes separately (ignoring the pair assignments), and the interpolation between the two. Note that P-LOOP imputing potential differences most closely matches the Regression 2 method, as both methods account for pairing and use the differences and averages of the covariates for making adjustments.

\begin{table}[H] 
	\caption{\label{tab02}Comparison of Methods}
	\centering
	\begin{tabular}{l|rr|rr}
		\toprule
		& \multicolumn{2}{c |}{\textbf{Pretest}} & \multicolumn{2}{c}{\textbf{Pretest and School Type}}\\
		\textbf{Method} & \textbf{Point Est.} & \textbf{Nominal Var.} & \textbf{Point Est.} & \textbf{Nominal Var.}\\ 
		\midrule
		Simple Difference & -6.82 & 9.82 & -6.82 & 9.82\\
		Regression 1 & -2.61 & 6.18 & -2.61 & 6.18\\
		Regression 2 & -2.60 & 6.56 & -2.27 & 4.57\\ 
		P-LOOP (differences) & -2.79 & 6.54 & -2.17 & 4.35\\ 
		P-LOOP (outcomes) & -2.04 & 5.55 & -1.81 & 4.02\\ 
		P-LOOP (interpolated) & -2.00 & 5.72 & -2.06 & 3.91\\ 
		\bottomrule
	\end{tabular}
\end{table}
Both P-LOOP and the methods of \textcite{fogarty2018regression} outperform the unadjusted estimator in terms of nominal variance. It is not clear ahead of time which regression method will perform better. Regression 1 outperforms Regression 2 when the pretest score is the only covariate, but Regression 2 outperforms Regression 1 when the school type is included. Note that the regression methods always account for the pair assignments. For the P-LOOP estimator, we see that it is better to impute the potential outcomes separately, and that the interpolation method imputes values closer to the potential outcomes imputation. With the interpolation method, we do not lose out on the precision gains from ignoring the pairs in our imputation, but we are still protected against a potential Simpson's paradox.
\section{Discussion} \label{discussion}
In paired experiments, the design of the experiment helps to enforce covariate balance between the treatment and control groups. While this design is often effective, it can be useful to make covariate adjustments to further improve precision. Covariate adjustments in paired experiments share many of the issues in completely randomized experiments; for example, it can be unclear ahead of time which covariates to use. A unique issue to paired experiments is the pair inclusion trade-off, so we must take particular care when making adjustments in paired experiments. Failing to account for the pair assignments can harm performance (for example, when a Simpson's paradox occurs), while including the paired structure when the pair assignments are not predictive can needlessly inflate variance. 

We present a design-based method for paired experiments, the P-LOOP estimator. To the best of our knowledge, this method is the first to directly address the pair inclusion trade-off. Generally, other methods account for the pairing, which protects against Simpson's paradox and other situations where the within pair trends differ from the overall trend. However, our method imputes two sets of potential outcomes, one excluding and one including the pair assignments, and automatically interpolates between the two. As we see in the Texas Schools data, this allows for improved precision. The P-LOOP estimator is also the first method for paired experiments that involves sample splitting and the use of machine learning methods to impute potential outcomes, building on the flexible approaches used in completely randomized experiments (\textcite{Aronow}, \textcite{Wager}, \textcite{chernozhukov2018double}, \textcite{wu2018loop}, \textcite{spiess2017optimal}, and \textcite{rotheflexible}). This flexibility can be beneficial in several ways, such as allowing for automatic variable selection or high dimensional covariates.

Finally, logical extensions to the P-LOOP estimator include block randomized experiments and experiments with multiple treatments. As with paired experiments, it can be unclear whether to include the block assignments when making covariate adjustments. However, while paired experiments can be treated essentially as Bernoulli randomized experiments, this is not the case for blocked experiments and the variance estimation procedure outlined in this paper would necessarily be modified.

\section{Acknowledgments}
We would like to thank Zhenke Wu for helpful comments and suggestions. We would also like to thank John Pane and Adam Sales for providing the data set used in Section \ref{schoolsdata}.

\printbibliography

\appendix
\section{True Variance of the P-LOOP Estimator} \label{truevar}
First, we calculate the variance of a single $\tauhi$:
\begin{align*}
\Var(\hat{\tau}_i) &= \E\left[\Var(\tauhi|\dhi)\right] + \Var\left[\E(\tauhi|\dhi)\right] \\
&= \E\left[\Var(\tauhi|\dhi)\right] + \Var(\tau_i) \\
&= \E\bsq{\Var\bparens{(W_i - \dhi)T_i + (W_i + \dhi)(1-T_i)|\dhi}} \\
&= \E\bsq{\Var\bparens{(a_i - \dhi)T_i + (b_i + \dhi)(1-T_i)|\dhi}} \\
&= \E\bsq{\Var\bparens{(a_i - b_i - 2\dhi)T_i + b_i + \dhi|\dhi}} \\
&= \E\bsq{(2d_i - 2\dhi)^2\Var\bparens{T_i|\dhi}} \\
&= \E\bsq{4(d_i - \dhi)^2 \times 1/4} \\
&= \E\bsq{(d_i - \dhi)^2} = \text{MSE}(\dhi).
\end{align*}
Let $\gamma_{ij} = \text{Cov}(\tauhi,\tauh{j})$. Then we have the following expression for the variance of the LOOP estimator
\begin{align*}
\Var(\hat{\tau}) &=  \Var\bparens{\frac{1}{N}\sum_{i = 1}^{N}\tauhi}\\
&= \frac{1}{N^2}\bparens{\sum_{i=1}^{N}\Var(\tauhi)
	+
	\sum_{i\ne j}\text{Cov}(\tauhi,\tauh{j})} \\
&=  \frac{1}{N^2}\bparens{\sum_{i=1}^{N}\text{MSE}(\dhi)
	+
	\sum_{i\ne j}\gamma_{ij}}. 
\end{align*}
\section{Handling the Covariance Terms $\gamma_{ij}$} \label{negligibility}
In this section, we further discuss the covariance terms $\gamma_{ij}$. First define $U_i = 2T_i - 1$; that is, $U_i = 1$ if $T_i = 1$, and $U_i = -1$ if $T_i = 0$. Note that $U_i$ has expectation 0.  Then we can rewrite $\tauhi$ as $W_i - \dhi U_i$. We have the following expression for $\gamma_{ij}$:

\begin{align*}
\gamma_{ij}
&= \Cov(\tauhi,\tauh{j}) \\
&= \Cov(W_i - \dhi U_i, W_j - \dhj U_j) \\
&= \Cov(W_i,W_j) - \Cov(W_i, \dhj U_j) - \Cov(\dhi U_i, W_j) + \Cov(\dhi U_i, \dhj U_j) \\
\end{align*}
The first term is zero, as $W_i$ and $W_j$ are independent due to the independence of $T_i$ and $T_j$. The second and third terms are also zero. Note that $U_j$ is independent of $W_i$ due to the independence of $T_i$ and $T_j$, and recall that $\dhj$ is independent of $T_j$ (and therefore $U_j$). Then we have for the second term
\begin{align*}
\Cov(W_i, \dhj U_j) 
&= \E(W_i \dhj U_j) - \E(W_i)\E(\dhj U_j) \\
&= \E(W_i \dhj) \E(U_j) - \E(W_i)\E(\dhj) \E(U_j) \\
&= 0
\end{align*}
where the last line follows because $\E(U_j) = 0$. We therefore have
\begin{align*}
\gamma_{ij} = \Cov(\dhi U_i, \dhj U_j). 
\end{align*}
Using a procedure outlined by \textcite{wu2018loop}, we can obtain an unbiased estimate of $\gamma_{ij}$. This procedure involves leaving out two pairs (rather than a single pair) at a time and is therefore computationally expensive. However, estimating these terms is often unnecessary, as they are generally negligible by an identical argument to one presented in \textcite{wu2018loop}. Under the setup and notation of Section \ref{notation} for Bernoulli randomized experiments, define an estimate the individual treatment effect for observation $i$ as:
\begin{align*}
\hat{\delta}_i &= 2(Y_i - \hat{m}_i)T_i - 2(Y_i - \hat{m}_i)(1-T_i)\\
&= 2(Y_i - \hat{m}_i)U_i.
\end{align*}
Note that we use $\hat\delta_i$ rather than $\hat\tau_i$ to avoid confusion with the notation for paired experiments.
\textcite{wu2018loop} show that for any observations $i$ and $j$, the covariance of these individual treatment effect estimates is
\begin{align*}
\Cov(\hat{\delta}_i,\hat{\delta}_j) = \Cov(\mhi U_i, \hat{m}_j U_j)
\end{align*}
and note that these terms are generally negligible in the sense that for many estimators of practical interest, $\Cov(\mhi U_i, \hat{m}_j U_j)$ goes to zero faster than $1/N$. 

\section{Bound on the Mean Squared Error of $\dhi$} \label{varbound}
We bound the term $\frac{1}{N^2}\sum_{i=1}^{N}\text{MSE}(\dhi)$. We can express the mean squared error of $\dhi$ in terms of the mean squared errors of $\ahi$ and $\bhi$:
\begin{align*}
\text{MSE}(\dhi)
&= \bsq{\E(d_i - \dhi)}^2 + \Var(\dhi) \\
&= \bsq{\bparens{\frac{1}{2}\E(a_i - \ahi) - \frac{1}{2}\E(b_i - \bhi)}}^2 + \frac{1}{4}\bsq{\Var(\ahi) + \Var(\bhi) - 2\text{Cov}(\ahi,\bhi)} \\
&= \frac{1}{4}\bsq{\text{Bias}(\ahi)}^2 + \frac{1}{4}\bsq{\text{Bias}(\bhi)}^2 - \frac{1}{2}\text{Bias}(\ahi)\text{Bias}(\bhi) + \\
& \hspace{.75cm} \frac{1}{4}\bsq{\Var(\ahi) + \Var(\bhi) - 2\text{Cov}(\ahi,\bhi)} \\
&= \frac{1}{4}\text{MSE}(\ahi) + \frac{1}{4}\text{MSE}(\bhi) + \frac{1}{2}\bsq{-\text{Bias}(\ahi)\text{Bias}(\bhi) -\text{Cov}(\ahi,\bhi)} \\
&\leq \frac{1}{4}\text{MSE}(\ahi) + \frac{1}{4}\text{MSE}(\bhi) + \frac{1}{2}\sqrt{\text{MSE}(\ahi)\text{MSE}(\bhi)}.
\end{align*}
This last inequality follows from the fact that $-\text{Bias}(\ahi)\text{Bias}(\bhi) -\text{Cov}(\ahi,\bhi) \leq \sqrt{\text{MSE}(\ahi)\text{MSE}(\bhi)}$. We then have the following bound:
\begin{align*}
\frac{1}{N^2}\sum_{i = 1}^N\text{MSE}(\dhi) 
&\leq \frac{1}{N^2}\sum_{i = 1}^N\bsq{\frac{1}{4}\text{MSE}(\ahi) + \frac{1}{4}\text{MSE}(\bhi) + \frac{1}{2}\sqrt{\text{MSE}(\ahi)\text{MSE}(\bhi)}} \\
&= \frac{1}{N}\bsq{\frac{1}{4N}\sum_{i = 1}^N\text{MSE}(\ahi) + \frac{1}{4N}\sum_{i = 1}^N\text{MSE}(\bhi) + \frac{1}{2}\sum_{i = 1}^N\frac{1}{N}\sqrt{\text{MSE}(\ahi)\text{MSE}(\bhi)}} \\
&\leq \frac{1}{N}\bsq{\frac{1}{4N}\sum_{i = 1}^N\text{MSE}(\ahi) + \frac{1}{4N}\sum_{i = 1}^N\text{MSE}(\bhi) + \frac{1}{2}\sqrt{\frac{1}{N}\sum_{i = 1}^N\text{MSE}(\ahi)\frac{1}{N}\sum_{i = 1}^N\text{MSE}(\bhi)}} \\
&= \frac{1}{N}\bparens{\frac{1}{4}M_a + \frac{1}{4}M_b + \frac{1}{2}\sqrt{M_aM_b}}
\end{align*}
where we define
\begin{align*}
M_a &= \frac{1}{N}\sum_{i = 1}^N\text{MSE}(\ahi) \\
M_b &= \frac{1}{N}\sum_{i = 1}^N\text{MSE}(\bhi).
\end{align*}
\end{document}